\documentclass[reprint,nofootinbib,showpacs]{revtex4}
\usepackage{graphicx}  
\usepackage{dcolumn}   
\usepackage{bm}        
\usepackage{amssymb}
\usepackage{hyperref}
\begin{document}   

\title{\large\bf Phenomenology of $\Xi_b \to \Xi_c\,\tau\,\nu$ decays}
\author{Rupak~Dutta${}$}
\email{rupak@phy.nits.ac.in}   
\affiliation{
National Institute of Technology Silchar, Silchar 788010, India\\
}

\begin{abstract}
Deviations from the standard model prediction have been reported in various semileptonic $B$ decays mediated via $b \to c$ charged current
interactions. In this context, we analyze semileptonic baryon decays $\Xi_b \to \Xi_c\,\tau\,\nu$ using the helicity formalism. We
report numerical results on various observables such as the decay rate, ratio of branching ratio, lepton side forward backward asymmetries, 
longitudinal polarization fraction of the lepton, and the convexity parameter for this decay mode using results of relativistic quark model.
We also provide an estimate of the new physics effect on these observables under various new physics scenarios. 
\end{abstract}


\maketitle

\section{Introduction}
\label{int}
Lepton flavor universality~(LFU) violation has been reported in various semileptonic $B$ meson decays mediated via $b \to c$ charged current
and $b \to s\,l^+\,l^-$ neutral current interactions. A combined excess of about $4.1\sigma$ from the standard model~(SM) prediction has been
reported by HFLAV~\cite{hflav} for $R_D$ and $R_{D^{\ast}}$, where $R_{D^(\ast)}$ represents the ratio of branching ratios of 
$B \to D^{(\ast)}\tau\nu$ to the corresponding $B \to D^{(\ast)}\,l\nu$ decays. Similarly, significant deviation from the SM expectation is 
observed in the LFU ratios $R_K^{(*)} = \mathcal B(B \to K^{(*)}\,\mu\mu)/\mathcal B(B \to K^{(*)}\,e\,e)$ mediated via flavor changing
neutral current~(FCNC) decays. 
Measurement of $R_K$ in the range $1< q^2 < 6\,{\rm GeV^2}$ deviates from the SM prediction at $2.6\sigma$ level~\cite{Aaij:2014ora}.
Similarly, the measured value of
$R_{K^{\ast}}$ in the dilepton invariant mass $q^2 = [0.045,\,1.1]\,{\rm GeV^2}$ and $[1.1,\,6.0]\,{\rm GeV^2}$~\cite{Aaij:2017vbb}
deviates from the SM prediction at approximately $2.1\sigma$ and $2.4\sigma$, respectively. 
Very recently, LHCb~\cite{Aaij:2017tyk} has measured the value of the ratio of branching ratio 
$R_{J/\Psi}=\mathcal B(B_c \to J/\Psi\,\tau\nu)/\mathcal B(B_c \to J/\Psi\,\mu\nu)$ to be $0.71\pm 0.17\pm 0.18$.
Comparing this measured value with the SM prediction~\cite{Wen-Fei:2013uea,Dutta:2017xmj, Ivanov:2005fd}, we find 
the discrepancy to be more than $2\sigma$.

Inspired by the anomalies present in the meson decays mediated via $b \to c$ charged current interactions, we study the corresponding baryon 
decays $\Xi_b \to \Xi_c\,\tau\,\nu$ within the SM and within various NP scenarios using the $\Xi_b \to \Xi_c$ form factors obtained from 
relativistic quark model. The $\Xi_b \to \Xi_c\,l\,\nu$ decays has been studied by various authors~\cite{Ebert:2006rp, Singleton:1990ye, 
Cheng:1995fe, Ivanov:1996fj, Ivanov:1998ya, Cardarelli:1998tq, Albertus:2004wj, Korner:1994nh}. In this paper, we use an 
effective theory formalism in the presence of NP and give prediction on various observables such as the decay rate, ratio of branching ratio, 
lepton side forward backward asymmetries, longitudinal polarization fraction of the lepton, and the convexity parameter for the 
$\Xi_b \to \Xi_c\,\tau\,\nu$ decays. Earlier discussion, however, have not looked into the $\Xi_b \to \Xi_c\,\tau\,\nu$ decays. To analyze
the effect of NP couplings on various observables, we use $1\sigma$ constraints coming from the measured values of the ratio of branching
ratios $R_D$ and $R_{D^{\ast}}$. The constraint coming from $B_c$ meson decay width is also discussed in details. We, however, do not use the
constraint coming from the $R_{J/\Psi}$ measurement as the error associated with it rather large.

Our paper is organized as follows. In section.~\ref{ehha}, we start with the most general effective Lagrangian in the presence of NP for
the $b \to c\tau\nu$ decays that is valid at renormalization scale $\mu = m_b$. A brief discussion on $\Xi_b \to \Xi_c$ transition form
factors is also presented. In section.~\ref{ehha}, we write down the helicity amplitudes and we define several observables such as the 
decay rate, ratio of branching ratio, $\tau$ polarization fraction, forward backward asymmetries, and the convexity parameter for the
$\Xi_b \to \Xi_c\,l\,\nu$ decays. In section.~\ref{rd}, we present our numerical results for all the observables defined in 
section.~\ref{ehha}. Finally, we present a brief summary of our results and conclude in section.~\ref{con}.
\section{Effective Lagrangian, heavy baryon form factors, and helicity amplitudes}
\label{ehha}
\subsection{Effective weak Lagrangian}
\label{ewl}
In the presence of NP, the effective weak Lagrangian for the $b \to c\,l\,\nu$ transition decays valid at renormalization scale
$\mu = m_b$ can be written as~\cite{Bhattacharya, Cirigliano}
\begin{eqnarray}
\mathcal L_{\rm eff} &=&
-\frac{4\,G_F}{\sqrt{2}}\,V_{cb}\,\Bigg\{(1 + V_L)\,\bar{l}_L\,\gamma_{\mu}\,\nu_L\,\bar{c}_L\,\gamma^{\mu}\,b_L +
V_R\,\bar{l}_L\,\gamma_{\mu}\,\nu_L\,\bar{c}_R\,\gamma^{\mu}\,b_R 
+
\widetilde{V}_L\,\bar{l}_R\,\gamma_{\mu}\,\nu_R\,\bar{c}_L\,\gamma^{\mu}\,b_L +
\widetilde{V}_R\,\bar{l}_R\,\gamma_{\mu}\,\nu_R\,\bar{c}_R\,\gamma^{\mu}\,b_R \nonumber \\
&&+
S_L\,\bar{l}_R\,\nu_L\,\bar{c}_R\,b_L +
S_R\,\bar{l}_R\,\nu_L\,\bar{c}_L\,b_R 
+
\widetilde{S}_L\,\bar{l}_L\,\nu_R\,\bar{c}_R\,b_L +
\widetilde{S}_R\,\bar{l}_L\,\nu_R\,\bar{c}_L\,b_R 
+ 
T_L\,\bar{l}_R\,\sigma_{\mu\nu}\,\nu_L\,\bar{c}_R\,\sigma^{\mu\nu}\,b_L \nonumber \\
&&+
\widetilde{T}_L\,\bar{l}_L\,\sigma_{\mu\nu}\,\nu_R\,\bar{c}_L\,\sigma^{\mu\nu}\,b_R\Bigg\} + {\rm H.c.}\,,
\end{eqnarray}
where $G_F$ is the Fermi constant, $ V_{cb}$ is the relevant Cabibbo-Kobayashi-Maskawa~(CKM) Matrix element, and $(c,\,b,\,l,\,\nu)_{R,\,L} = 
\Big(\frac{1 \pm \gamma_5}{2}\Big)\,(c,\,b,\,l,\,\nu)$. The NP couplings, associated with new vector, scalar, and tensor interactions, 
denoted by $V_{L,R}$, $S_{L,R}$, and $ T_{L}$ involve
left-handed neutrinos, whereas, the NP couplings denoted by $\widetilde{V}_{L,R}$, $\widetilde{S}_{L,R}$, and $\widetilde{T}_{L}$ involve 
right-handed neutrinos.
We consider NP contributions coming from vector and scalar type of interactions only. We neglect the contributions coming from NP couplings 
that involves right-handed neutrinos, i.e,
$\widetilde{V}_{L,R} = \widetilde{S}_{L,R}$ = $\widetilde{T}_{L} = 0$.  All the NP couplings are assumed to be real for our analysis. With 
these assumptions, we obtain
\begin{eqnarray}
\label{leff}
\mathcal L_{\rm eff} &=&
-\frac{G_F}{\sqrt{2}}\,V_{cb}\,\Bigg\{G_V\,\bar{l}\,\gamma_{\mu}\,(1 - \gamma_5)\,\nu_l\,\bar{c}\,\gamma^{\mu}\,b -
G_A\,\bar{l}\,\gamma_{\mu}\,(1 - \gamma_5)\,\nu_l\,\bar{c}\,\gamma^{\mu}\,\gamma_5\,b 
+
G_S\,\bar{l}\,(1 - \gamma_5)\,\nu_l\,\bar{c}\,b - G_P\,\bar{l}\,(1 - \gamma_5)\,\nu_l\,\bar{c}\,\gamma_5\,b \Bigg\} \nonumber \\
&&+ 
{\rm H.c.}\,,
\end{eqnarray}
where 
\begin{eqnarray*} 
&&G_V = 1 + V_L + V_R\,,\qquad\qquad
G_A = 1 + V_L - V_R\,, \qquad\qquad
G_S = S_L + S_R\,,\qquad\qquad
G_P = S_L - S_R\,.
\end{eqnarray*}
The SM contribution can be obtained once we set $V_{L,R} = S_{L,R} = 0$ in Eq.~(\ref{leff}).
\subsection{$\Xi_b \to \Xi_c$ transition form factors}
\label{ffs}
The hadronic matrix elements of vector and axial vector currents between two spin half baryons are parametrized in terms of various hadronic
form factors as follows
\begin{eqnarray}
&&M_{\mu}^V = \langle B_2,\lambda_2 | J_{\mu}^V | B_1,\lambda_1\rangle = \bar{u}_2(p_2,\lambda_2)\Big[f_1(q^2)\gamma_{\mu} + if_2(q^2)
\sigma_{\mu\nu}\,q^{\nu} + f_3(q^2)q_{\mu}\Big]u_1(p_1,\lambda_1)\,,\nonumber \\
&&M_{\mu}^A= \langle B_2,\lambda_2 | J_{\mu}^A | B_1,\lambda_1\rangle = \bar{u}_2(p_2,\lambda_2)\Big[g_1(q^2)\gamma_{\mu} + ig_2(q^2)
\sigma_{\mu\nu}\,q^{\nu} + g_3(q^2)q_{\mu}\Big]\,\gamma_5\,u_1(p_1,\lambda_1)\,,
\end{eqnarray}
where $q^{\mu} = (p_1 - p_2)^{\mu}$ is the four momentum transfer, $\lambda_1$ and $\lambda_2$ are the helicities of the parent and daughter 
baryons, respectively and $\sigma_{\mu\nu} = \frac{i}{2}\,[\gamma_{\mu}, \gamma_{\nu}]$. Here $B_1$ represents $\Xi_b$ baryon and $B_2$ 
represents $\Xi_c$ baryon, respectively. When both baryons are heavy, it is also convenient to parmetrize the matrix element in terms of the
four velocities $v^{\mu}$ and $v^{{\prime}^{\mu}}$ as follows:
\begin{eqnarray}
&&M_{\mu}^V= \langle B_2,\lambda_2 | J_{\mu}^V | B_1,\lambda_1\rangle = \bar{u}_2(p_2,\lambda_2)\Big[F_1(\omega)\gamma_{\mu} + F_2(\omega)
\,v_{\mu} + F_3(\omega)v^{\prime}_{\mu}\Big]u_1(p_1,\lambda_1)\,,\nonumber \\
&&M_{\mu}^A= \langle B_2,\lambda_2 | J_{\mu}^A | B_1,\lambda_1\rangle = \bar{u}_2(p_2,\lambda_2)\Big[G_1(\omega)\gamma_{\mu} + G_2(\omega)
\,v_{\mu} + G_3(\omega)v^{\prime}_{\mu}\Big]\gamma_5\,u_1(p_1,\lambda_1)\,,
\end{eqnarray}
where $\omega = v\cdot v^{\prime} = (m_{B_1}^2+m_{B_2}^2 - q^2)/(2\,m_{B_1}\,m_{B_2})$ and $m_{B_1}$ and $m_{B_2}$ are the masses of $B_1$ 
and $B_2$ baryons, respectively. The two sets of form factors are related via
\begin{eqnarray}
&&f_1(q^2)=F_1(q^2) + (m_{B_2} + m_{B_1})\Big[\frac{F_2(q^2)}{2\,m_{B_1}} + \frac{F_3(q^2)}{2\,m_{B_2}}\Big]\,, \nonumber \\
&&f_2(q^2)=\frac{F_2(q^2)}{2\,m_{B_1}} + \frac{F_3(q^2)}{2\,m_{B_2}} \,, \nonumber \\
&&f_3(q^2)=\frac{F_2(q^2)}{2\,m_{B_1}} - \frac{F_3(q^2)}{2\,m_{B_2}} \,, \nonumber \\ 
&&g_1(q^2)=G_1(q^2) - (m_{B_2} - m_{B_1})\Big[\frac{G_2(q^2)}{2\,m_{B_1}} + \frac{G_3(q^2)}{2\,m_{B_2}}\Big]\,, \nonumber \\
&&g_2(q^2)=\frac{G_2(q^2)}{2\,m_{B_1}} + \frac{G_3(q^2)}{2\,m_{B_2}} \,, \nonumber \\
&&g_3(q^2)=\frac{G_2(q^2)}{2\,m_{B_1}} - \frac{G_3(q^2)}{2\,m_{B_2}} \,,
\end{eqnarray}
We use the equation of motion to find the hadronic matrix elements of scalar and pseudoscalar 
currents. That is
\begin{eqnarray}
&&\langle B_2,\lambda_2 | \bar{c}\,b | B_1,\lambda_1\rangle = \bar{u}_2(p_2,\lambda_2)\Big[f_1(q^2)\,\frac{\not q}{m_b - m_c} +  
f_3(q^2)\,\frac{q^2}{m_b - m_c} \Big]u_1(p_1,\lambda_1)\,,\nonumber \\
&&\langle B_2,\lambda_2 |  \bar{c}\,\gamma_5\,b  | B_1,\lambda_1\rangle = \bar{u}_2(p_2,\lambda_2)\Big[-g_1(q^2)\frac{\not q}
{m_b + m_c} -  g_3(q^2)\,\frac{q^2}{m_b + m_c} \Big]\,\gamma_5\,u_1(p_1,\lambda_1)\,,
\end{eqnarray}
where $m_b$ is the mass of $b$ quark and $m_c$ is the mass of $c$ quark evaluated at renormalization scale 
$\mu = m_b$, respectively. 
For the various invariant form factors $F_i$'s and $G_i$'s, we follow Ref~.\cite{Ebert:2006rp}. The relevant equations pertinent for our 
calculation are as follows:
\begin{eqnarray}
&&F_1(\omega)=\zeta(\omega) + \Big(\frac{\bar{\Lambda}}{2\,m_b} + \frac{\bar{\Lambda}}{2\,m_c}\Big)\Big[2\chi(\omega) + \zeta(\omega)\Big]\,,
\nonumber \\
&&G_1(\omega)=\zeta(\omega) + \Big(\frac{\bar{\Lambda}}{2\,m_b} + \frac{\bar{\Lambda}}{2\,m_c}\Big)\Big[2\chi(\omega) + 
\frac{\omega-1}{\omega+1}\,\zeta(\omega)\Big]\,,\nonumber \\
&&F_2(\omega)=G_2(\omega)=-\frac{\bar{\Lambda}}{2\,m_c}\,\frac{2}{\omega+1}\,\zeta(\omega)\,,\nonumber \\
&&F_3(\omega)=-G_3(\omega)=-\frac{\bar{\Lambda}}{2\,m_b}\,\frac{2}{\omega+1}\,\zeta(\omega)\,,
\end{eqnarray}
where $\bar{\Lambda}=m_{B_1} - m_b$ is the difference of the baryon and the heavy quark mass in the heavy quark limit $m_b \to \infty$. Here
$\zeta(\omega)$ denotes the Isgur-Wise function. The additional function $\chi(\omega)$ appears due to the $1/m_b$ correction to the
heavy quark effective theory~(HQET) Lagrangian. Near
the zero recoil point of the final baryon $\omega = 1$, the functions $\zeta(\omega)$ and $\chi(\omega)$ can be expressed as
\begin{eqnarray}
&&\zeta(\omega) = 1 - \rho_{\zeta}^2\,(\omega -1) + c_{\zeta}(\omega -1)^2\,,\nonumber \\
&&\chi(\omega)=\rho_{\chi}^2\,(\omega -1) + c_{\chi}(\omega -1)^2\,,
\end{eqnarray}
where $\rho_{\zeta}^2$ and $c_{\zeta}$ represent the slope and the curvature of the Isgur-Wise functions, respectively. We refer to
Ref.~\cite{Ebert:2006rp} for all the omitted details.
\subsection{Helicity amplitudes}
We now proceed to discuss the helicity amplitudes for baryonic $b \to c\,l\,\nu$ decay mode. The helicity amplitudes can be defined 
by~\cite{Korner,Gutsche:2014zna,Gutsche:2015mxa}
\begin{eqnarray}
&&H_{\lambda_2\,\lambda_W}^{V/A} = M_{\mu}^{V/A}(\lambda_2)\,\epsilon^{\dagger^{\mu}}(\lambda_W)\,,
\end{eqnarray}
where $\lambda_2$ and $\lambda_W$ denote the helicities of the daughter baryon and $W^-_{\rm off - shell}$, respectively. The total 
left - chiral helicity amplitude can be written as
\begin{eqnarray}
&&H_{\lambda_2\,\lambda_W} = H_{\lambda_2\,\lambda_W}^{V} - H_{\lambda_2\,\lambda_W}^{A}\,. 
\end{eqnarray}
In terms of the various form factors and the NP couplings, the helicity amplitudes can be written as~\cite{Shivashankara:2015cta,
Dutta:2015ueb}
\begin{eqnarray}
&&H_{\frac{1}{2}\,0}^V = G_V\,\frac{\sqrt{Q_-}}{\sqrt{q^2}}\,\Big[(m_{B_1} + m_{B_2})\,f_1(q^2) - q^2\,f_2(q^2)\Big]\,, \nonumber \\
&&H_{\frac{1}{2}\,0}^A = G_A\,\frac{\sqrt{Q_+}}{\sqrt{q^2}}\,\Big[(m_{B_1} - m_{B_2})\,g_1(q^2) + q^2\,g_2(q^2)\Big]\,, \nonumber \\
&&H_{\frac{1}{2}\,1}^V = G_V\,\sqrt{2\,Q_-}\,\Big[-f_1(q^2) + (m_{B_1} + m_{B_2})\,f_2(q^2)\Big]\,, \nonumber \\
&&H_{\frac{1}{2}\,1}^A = G_A\,\sqrt{2\,Q_+}\,\Big[-g_1(q^2) - (m_{B_1} - m_{B_2})\,g_2(q^2)\Big]\,, \nonumber \\
&&H_{\frac{1}{2}\,t}^V = G_V\,\frac{\sqrt{Q_+}}{\sqrt{q^2}}\,\Big[(m_{B_1} - m_{B_2})\,f_1(q^2) + q^2\,f_3(q^2)\Big]\,, \nonumber \\
&&H_{\frac{1}{2}\,t}^A = G_A\,\frac{\sqrt{Q_-}}{\sqrt{q^2}}\,\Big[(m_{B_1} + m_{B_2})\,g_1(q^2) - q^2\,g_3(q^2)\Big]\,, 
\end{eqnarray}
where $Q_{\pm} = (m_{B_1} \pm m_{B_2})^2 - q^2$. Either from parity or from explicit calculation, one can show that 
$H^V_{-\lambda_2\, -\lambda_W} = H^V_{\lambda_2\, \lambda_W}$ and $H^A_{-\lambda_2\, -\lambda_W} = - H^A_{\lambda_2\, \lambda_W}$. 
Similarly, the scalar and pseudoscalar helicity amplitudes associated with the NP couplings $G_S$ and $G_P$ can be written 
as~\cite{Shivashankara:2015cta,Dutta:2015ueb}
\begin{eqnarray}
&&H_{\frac{1}{2}\,0}^{SP} = H_{\frac{1}{2}\,0}^{S} - H_{\frac{1}{2}\,0}^{P}\,, \nonumber \\
&&H_{\frac{1}{2}\,0}^{S} = G_S\,\frac{\sqrt{Q_+}}{m_b - m_{q^{\prime}}}\,\Big[(m_{B_1} - m_{B_2})\,f_1(q^2) + q^2\,f_3(q^2)\Big]\,, 
\nonumber \\
&&H_{\frac{1}{2}\,0}^{P} = G_P\,\frac{\sqrt{Q_-}}{m_b + m_{q^{\prime}}}\,\Big[(m_{B_1} + m_{B_2})\,g_1(q^2) - q^2\,g_3(q^2)\Big]\,.
\end{eqnarray}
Moreover,  we have $H_{\lambda_2\,\lambda_{\rm NP}}^S = H_{-\lambda_2\,-\lambda_{\rm NP}}^S$ and 
$H_{\lambda_2\,\lambda_{\rm NP}}^P = -H_{-\lambda_2\,-\lambda_{\rm NP}}^P$ from parity argument or from explicit calculation.

We follow Ref.~\cite{Shivashankara:2015cta,Dutta:2015ueb} and write the differential angular distribution for the three body 
$B_1 \to B_2\,l\,\nu$ decays in the presence of NP as
\begin{eqnarray}
\label{dtdq2dcth}
&&\frac{d\Gamma(B_1 \to B_2\,l\,\nu)}{dq^2\,d\cos\theta_l}= N\,\Big(1-\frac{m_l^2}{q^2}\Big)^2\Big[A_1 + \frac{m_l^2}{q^2}\,A_2 + 2\,A_3 + 
\frac{4\,m_l}{\sqrt{q^2}}\,A_4\Big]\,,
\end{eqnarray}
where
\begin{eqnarray}
N &=& \frac{G_F^2\,|V_{q^{\prime}\,b}|^2\,q^2\,|\vec{p}_{B_2}|}{512\,\pi^3\,m_{B_1}^2}\,,
\nonumber \\
A_1 &=& 2\,\sin^2\theta_l\,\Big(H_{\frac{1}{2}\,0}^2 + H_{-\frac{1}{2}\,0}^2\Big) +\Big(1 - \cos\theta_l\Big)^2\,H_{\frac{1}{2}\,1}^2 + 
\Big(1 + \cos\theta_l\Big)^2\,H_{-\frac{1}{2}\,-1}^2\,, \nonumber \\
A_2 &=& 2\,\cos^2\theta_l\,\Big(H_{\frac{1}{2}\,0}^2 + H_{-\frac{1}{2}\,0}^2\Big) + \sin^2\theta_l\,\Big(H_{\frac{1}{2}\,1}^2 + 
H_{-\frac{1}{2}\,-1}^2\Big) + 2\,\Big(H_{\frac{1}{2}\,t}^2 + H_{-\frac{1}{2}\,t}^2\Big) - \nonumber \\
&&4\,\cos\theta_l\,\Big(H_{\frac{1}{2}\,t}\,H_{\frac{1}{2}\,0} + H_{-\frac{1}{2}\,t}\,H_{-\frac{1}{2}\,0}\Big)\,, \nonumber \\
A_3 &=& (H^{SP}_{\frac{1}{2}\,0})^2 + (H^{SP}_{-\frac{1}{2}\,0})^2\,, \nonumber \\
A_4 &=& -\cos\theta_l\,\Big(H_{\frac{1}{2}\,0}\,H^{SP}_{\frac{1}{2}\,0} + H_{-\frac{1}{2}\,0}\,H^{SP}_{-\frac{1}{2}\,0}\Big) + 
\Big(H_{\frac{1}{2}\,t}\,H^{SP}_{\frac{1}{2}\,0} + H_{-\frac{1}{2}\,t}\,H^{SP}_{-\frac{1}{2}\,0}\Big)\,.
\end{eqnarray}
Here $ |\vec{p}_{B_2}| = \sqrt{\lambda(m_{B_1}^2,\,m_{B_2}^2,\,q^2)}/2\,m_{B_1}$ is the momentum of the outgoing baryon $B_2$, 
where $\lambda(a,\,b,\,c) = a^2 + b^2 + c^2 - 2\,(a\,b + b\,c + c\,a)$. We denote $\theta_l$ as the angle between the daughter baryon 
$B_2$ and the lepton three momentum vector in the $q^2$ rest frame.
The differential decay rate can be obtained by integrating out $\cos\theta_l$ from Eq.~(\ref{dtdq2dcth}), i.e,
\begin{eqnarray}
&&\frac{d\Gamma(B_1 \to B_2\,l\,\nu)}{dq^2}= \frac{8\,N}{3}\,\Big(1-\frac{m_l^2}{q^2}\Big)^2\Big[B_1 + \frac{m_l^2}{2\,q^2}\,B_2 + 
\frac{3}{2}\,B_3 + \frac{3\,m_l}{\sqrt{q^2}}\,B_4\Big]\,,
\end{eqnarray}
where
\begin{eqnarray}
&&B_1 = H_{\frac{1}{2}\,0}^2 + H_{-\frac{1}{2}\,0}^2 + H_{\frac{1}{2}\,1}^2 + H_{-\frac{1}{2}\,-1}^2\,, \nonumber \\
&&B_2 = H_{\frac{1}{2}\,0}^2 + H_{-\frac{1}{2}\,0}^2 + H_{\frac{1}{2}\,1}^2 + H_{-\frac{1}{2}\,-1}^2 + 3\,\Big(H_{\frac{1}{2}\,t}^2 + 
H_{-\frac{1}{2}\,t}^2\Big)\,, \nonumber \\
&&B_3 = (H^{SP}_{\frac{1}{2}\,0})^2 + (H^{SP}_{-\frac{1}{2}\,0})^2\,, \nonumber \\
&&B_4 = H_{\frac{1}{2}\,t}\,H^{SP}_{\frac{1}{2}\,0} + H_{-\frac{1}{2}\,t}\,H^{SP}_{-\frac{1}{2}\,0}\,.
\end{eqnarray}
The ratio of branching ratios $R_{\Xi_c}$ is defined as
\begin{eqnarray}
\label{rdds}
&& R_{\Xi_c}=\frac{\mathcal B(\Xi_b \to \Xi_c\tau^- \bar{\nu}_\tau)}{\mathcal B( \Xi_b \to \Xi_c\, l^- \bar{\nu}_l)} \,,
\end{eqnarray}
where $l$ is either an electron or a muon.
We have also defined several $q^2$ dependent observables such as differential branching fractions ${\rm DBR}(q^2)$, ratio of branching 
fractions $R(q^2)$, forward backward asymmetries $A^l_{\rm FB}(q^2)$, the convexity parameter $C_F^l(q^2)$, and the longitudinal polarization
fraction of the lepton $P_l(q^2)$ for the baryonic $\Xi_b \to \Xi_c\,l\,\nu$ decay mode. Those are
\begin{eqnarray}
&&{\rm DBR}(q^2) = \Big(\frac{d\Gamma}{dq^2}\Big)\Big/\Gamma_{\rm tot}\,, \qquad\qquad
R(q^2)=\frac{DBR(q^2)\Big(B_1 \to B_2\,\tau\,\nu\Big)}{DBR(q^2)\Big(B_1 \to B_2\,l\,\nu\Big)}\,, \qquad\qquad
P_{l}(q^2)=\frac{d\Gamma(+)/dq^2 - d\Gamma(-)/dq^2}{d\Gamma(+)/dq^2 + d\Gamma(-)/dq^2}
\nonumber\\
&&A^l_{\rm FB}(q^2)=\Bigg\{\Big(\int_{-1}^{0}-\int_{0}^{1}\Big)d\cos \theta_l\frac{d\Gamma}{dq^2\,d\cos\theta_l}\Bigg\}\Big/
\frac{d\Gamma}{dq^2}\,, \qquad\qquad
C_F^l(q^2) = \frac{1}{\mathcal H_{\rm tot}}\,\frac{d^2\,W(\theta)}{d(\cos\theta)^2}\,,
\end{eqnarray}
where $d\Gamma(+)/dq^2$ and $d\Gamma(-)/dq^2$ denote the differential branching ratio of positive and negative helicity leptons, respectively.
Again
\begin{eqnarray}
&&W(\theta) = \frac{3}{8}\,\Big[A_1 + \frac{m_l^2}{q^2}\,A_2 + 2\,A_3 + \frac{4\,m_l}{\sqrt{q^2}}\,A_4\Big]\,, \nonumber \\
&& \mathcal H_{\rm tot} = \int\,d(\cos\theta)\,W(\theta)\,, \nonumber \\
&&\frac{d^2\,W(\theta)}{d(\cos\theta)^2} = \frac{3}{4}\,\Big(1 - \frac{m_l^2}{q^2}\Big)\,\Big[H_{\frac{1}{2}\,1}^2 + 
H_{-\frac{1}{2}\,-1}^2 - 2\,\Big(H_{\frac{1}{2}\,0}^2 + H_{-\frac{1}{2}\,0}^2\Big)\Big]\,.
\end{eqnarray}
we also give our predictions for the average values of the forward-backward asymmetry of the charged lepton $<A^l_{FB}>$, the convexity 
parameter $<C_F^l>$, and the longitudinal polarization of the lepton $<P_l>$ which are calculated by separately integrating the 
numerators and denominators over $q^2$.

\section{Results and Discussion}
\label{rd}
For definiteness, we first present all the inputs that are pertinent for our calculation. For the quark, lepton, and the baryon masses, we 
use $m_b(m_b) = 4.18\,{\rm GeV}$, $m_c(m_b) = 0.91\,{\rm GeV}$, $m_e = 0.510998928 \times 10^{-3}\,{\rm GeV}$, $m_{\mu} = 0.1056583715\,
{\rm GeV}$, $m_{\tau} = 1.77682\,{\rm GeV}$, $m_{\Xi_b} = 5.7919\,{\rm GeV}$, $m_{\Xi_c} = 2.46787\,{\rm GeV}$~\cite{Patrignani:2016xqp}. 
For the mean life 
time of $\Xi_b$ baryon, we use $\tau_{\Xi_b} = 1.479\times 10^{-12}\,{\rm s}$~\cite{Patrignani:2016xqp}. For the CKM matrix element 
$|V_{cb}|$, we have used 
the value $|V_{cb}| = (40.9 \pm 1.1) \times 10^{-3}$~\cite{Patrignani:2016xqp}. The relevant parameters for the form factor calculation are 
given in Table.~\ref{tab1}. We have used $\pm 10\%$ uncertainty in each of these parameters. We also report the most important experimental 
input parameters $R_D$ and $R_{D^{\ast}}$ in Table.~\ref{tab2}. We use the average values of $R_D$ and $R_{D^{\ast}}$ for our analysis. 
For the errors, we added the statistical and systematic uncertainties in quadrature.
\begin{table}[htbp]
\centering
\begin{tabular}{|c|c|c|c|c|}
\hline
$\bar{\Lambda}({\rm GeV})$ & $\rho_{\zeta}^2$ & $c_{\zeta}$ & $\rho_{\chi}^2$ & $c_{\chi}$ \\[0.2cm]
\hline
\hline
$0.970$ & $2.27$ & $3.87$ & $0.045$ & $0.036$ \\[0.2cm]
\hline
\end{tabular}
\caption{Parameters for the Isgur-Wise functions of $\Xi_b \to \Xi_c$ form factors taken from Ref.~\cite{Ebert:2006rp}}
\label{tab1}
\end{table}
\begin{table}[htbp]
\centering
\begin{tabular}{|c|c|c|}
\hline
Experiments & $ R_{D^{\ast}} $ & $R_D$ \\[0.2cm]
\hline
\hline
BABAR &$0.332\pm 0.024\pm 0.018$ &$0.440 \pm 0.058 \pm 0.042 $ \\[0.2cm]
BELLE &$0.293 \pm 0.038 \pm 0.015$ &$0.375 \pm 0.064 \pm 0.026$ \\[0.2cm]
BELLE &$0.302 \pm 0.030 \pm 0.011 $ & \\[0.2cm]
LHCb &$0.336 \pm 0.027 \pm 0.030 $ & \\[0.2cm]
BELLE &$0.270 \pm 0.035^{+ 0.028}_{-0.025}$ & \\[0.2cm]
LHCb &$0.285 \pm 0.019 \pm 0.029 $ & \\[0.2cm]
\hline
\hline
AVERAGE &$0.304 \pm 0.013 \pm 0.007$ &$0.407 \pm 0.039 \pm 0.024 $ \\[0.2cm]
\hline
\end{tabular}
\caption{Current status of $R_D$ and $R_{D^{\ast}}$~\cite{hflav}.}
\label{tab2}
\end{table}

There are two major sources of uncertainties in the calculation of the decay amplitudes. It may come either from not so well known input 
parameters such as CKM matrix elements or from 
hadronic input parameters such as form factors and decay constants. In order to gauge the effect of these above mentioned uncertainties on 
various observables, we use a random number generator and perform a random scan over all the theoretical input parameters such as CKM matrix 
elements, form factors, and decay constants within $1\sigma$ of their central values.
The SM central values and the corresponding $1\sigma$ ranges of all the observables for the $\Xi_b \to \Xi_c\,l\,\nu$ decays are presented 
in Table.~\ref{tab3}. We notice that there
are considerable changes while going from $e$ to $\tau$ mode, including even a sign change in the forward backward asymmetry parameter
$<A_{FB}^l>$.
The central values reported in Table.~\ref{tab3} are obtained using the central values of all the input parameters whereas, to find the 
$1\sigma$ range of all the observables, we vary all the input parameters such as CKM matrix elements, the hadronic form factors, and the 
decay constants within $1\sigma$ from their central values. We, however, do not include the uncertainties coming from the quark mass, 
lepton mass, baryon mass, and the mean life time as these are not important for our analysis.  
\begin{table}
\begin{center}
\begin{tabular}{|c|c|c|}
\hline
Observables  & Central value & $1\sigma$ range\\
\hline
\hline
$\mathcal B(\Xi_b \to \Xi_c\,e\,\nu)\%$ & $9.22 $ & $[5.62,\,14.47]$  \\[0.2cm]
$\mathcal B(\Xi_b \to \Xi_c\,\tau\,\nu)\%$ & $2.35$ & $[1.84,\,2.99]$ \\[0.2cm]
$R_{\Xi_c}$  & $0.255$ & $[0.205,\,0.330]$ \\[0.2cm]
$<A^e_{FB}>$  & $0.163$ & $[0.146,\,0.185]$ \\[0.2cm]
$<A^{\tau}_{FB}>$  & $-0.042$ & $[-0.027,\,-0.056]$ \\[0.2cm]
$<C_{F}^e>$ &$-0.697$ & $[-0.572,\,-0.781]$ \\[0.2cm]
$<C_{F}^{\tau}>$ &$-0.103$ &$[-0.100,\,-0.106]$ \\[0.2cm]
$<P_{e}>$ &$-1.00$ & $-1.00$ \\[0.2cm]
$<P_{\tau}>$ &$-0.317$ &$[-0.298,\,-0.338]$  \\[0.2cm]
\hline
\end{tabular}
\end{center}
\caption{Prediction of various observables within the SM.}
\label{tab3}
\end{table}

\begin{figure}[htbp]
\begin{center}
\includegraphics[width=5.9cm,height=4.5cm]{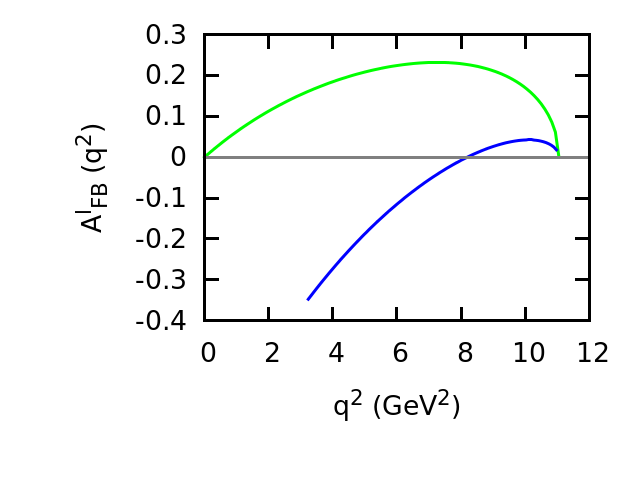}
\includegraphics[width=5.9cm,height=4.5cm]{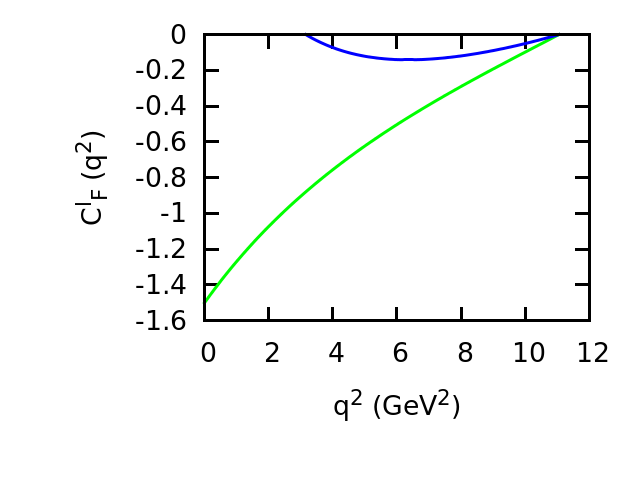}
\includegraphics[width=5.9cm,height=4.5cm]{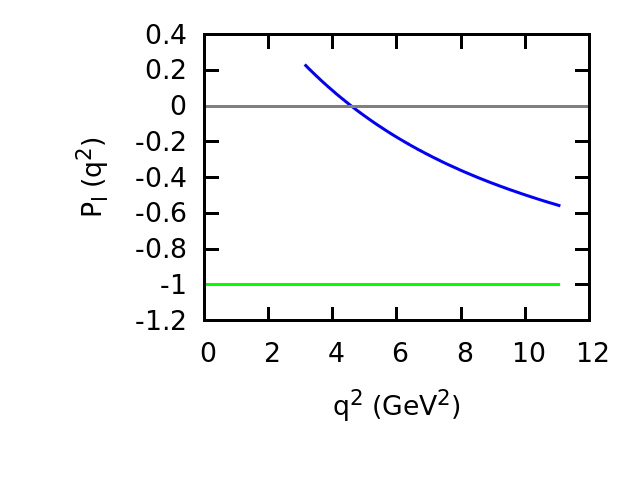}
\end{center}
\caption{$q^2$ dependence of $A^{l}_{FB}(q^2)$, $C_F^{l}(q^2)$, and $P_{l}(q^2)$ within the SM. 
Blue and green line corresponds to the $\tau$ and $e$ modes, respectively.}
\label{obs_sm}
\end{figure}

In Fig.~\ref{obs_sm}, we show the $q^2$ dependence of $A_{FB}^l(q^2)$, $C_F^l(q^2)$, and $P_l(q^2)$ within the SM for 
the $\tau$ and the $e$ modes. We observe
that the $q^2$ behavior of all the observables in the $e$ mode is quite different from the $\tau$ mode. 
The forward backward asymmetry parameter $A_{FB}^l(q^2)$ approaches zero at zero recoil for both the $e$ and the $\tau$ modes. 
We observe that although $A_{FB}^l(q^2)$ remains positive for the $e$ mode, it, 
however, becomes negative for the $\tau$ mode below $q^2 \equiv 8.0\,{\rm GeV^2}$. We observe a zero crossing in the $A_{FB}^l(q^2)$ parameter
for the $\tau$ mode. Similarly, for the $C_F^l(q^2)$ parameter, at zero recoil it approaches zero for both $e$ and $\tau$ modes. However, at
maximum recoil, $C_F^l(q^2)$ becomes zero for the $\tau$ mode, whereas, it becomes large and negative for the $e$ mode. Again, the convexity 
parameter remains very small in the whole $q^2$ region for the $\tau$ mode. The longitudinal polarization fraction of the charged lepton
$P_{l}(q^2)$ is $-1$ in the entire $q^2$ region for the $e$ mode. For the $\tau$ mode, we observe a zero crossing in the $P_{l}(q^2)$
parameter at $q^2 \equiv 5.0\,{\rm GeV^2}$ below which it becomes positive.

\begin{figure}[htbp]
\begin{center}
\includegraphics[width=7cm,height=6cm]{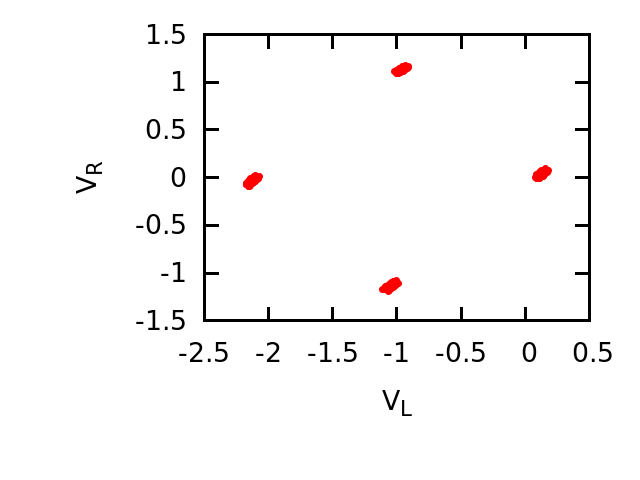}
\includegraphics[width=7cm,height=6cm]{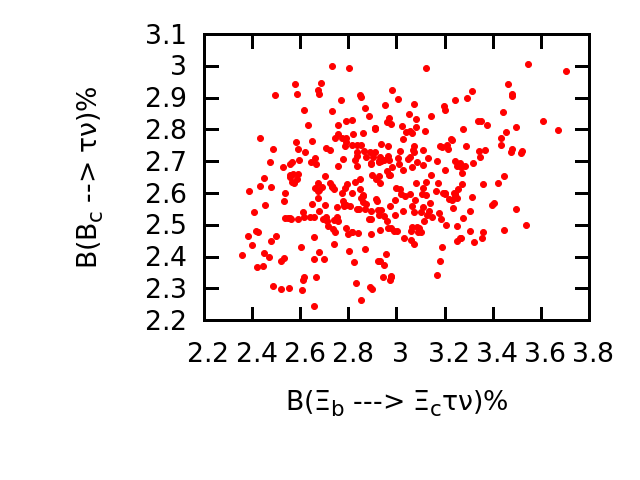}
\end{center}
\caption{Allowed regions of $V_L$ and $V_R$ obtained using the $1\sigma$ constraint coming from $R_D$ and $R_{D^{\ast}}$ are shown in the 
left panel and the corresponding ranges in $\mathcal B(B_c \to \tau\nu)$ and $\mathcal B(\Xi_b \to \Xi_c\tau\nu)$ in
the presence of these NP couplings are shown in the right panel.}
\label{vlvr}
\end{figure}

Now we proceed to discuss various NP scenarios. We want to see the effect of various NP couplings in a model independent way. In the first 
scenario, we assume that NP is coming from couplings associated with new vector type of interactions, i.e, from $V_L$ and $V_R$ only. We vary 
$V_L$ and $V_R$ while keeping $S_{L,\,R} = 0$. In order to determine the allowed NP parameter space, we impose $1\sigma$ constraint coming 
from the measured values of the ratio of branching
ratios $R_D$ and $R_{D^{\ast}}$. The allowed ranges in $V_L$ and $V_R$ that satisfies the $1\sigma$ experimental constraint are shown in 
the left panel of Fig.~\ref{vlvr}. In the right panel, we show the allowed ranges in $\mathcal B(B_c \to \tau\nu)$ and 
$\mathcal B(\Xi_b \to \Xi_c\tau\nu)$ obtained in this NP scenario. We see that $\mathcal B(B_c \to \tau\nu)$ obtained in this scenario is 
consistent with the $\mathcal B(B_c \to \tau\nu) \le 5\%$ obtained in the SM. The corresponding ranges of all the observables are listed in 
Table.~\ref{tab4}.
\begin{table}
\begin{center}
\begin{tabular}{|c|c|c|c|c|}
\hline
$\mathcal B(\Xi_b \to \Xi_c\tau\nu)\%$ & $R_{\Xi_c}$ & $<A_{FB}^{\tau}>$ & $<C_F^{\tau}>$ & $<P_{\tau}>$ \\[0.3cm]
\hline
\hline
$[2.36,\,3.71]$ & $[0.254,\,0.410]$ & $[-0.028,\,-0.346]$ & $[-0.101,\,-0.113]$ & $[-0.246,\,-0.333]$\\ [0.2cm]
\hline
\end{tabular}
\end{center}
\caption{Ranges of various observables with $V_L$ and $V_R$ NP couplings.}
\label{tab4}
\end{table}
We see a significant deviation from the SM prediction. Depending on the NP couplings $V_L$ and $V_R$, value of the observables can be either 
smaller or larger than the SM prediction. 

\begin{figure}[htbp]
\begin{center}
\includegraphics[width=5.9cm,height=4.5cm]{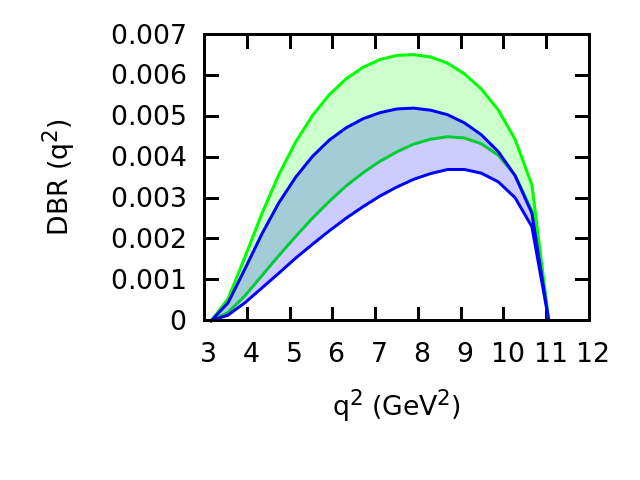}
\includegraphics[width=5.9cm,height=4.5cm]{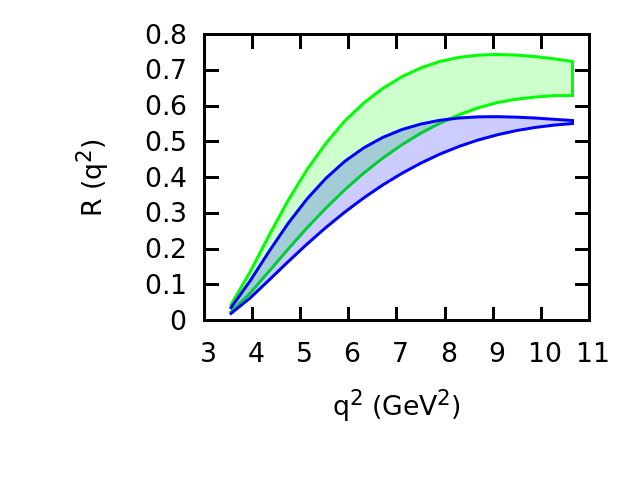}
\includegraphics[width=5.9cm,height=4.5cm]{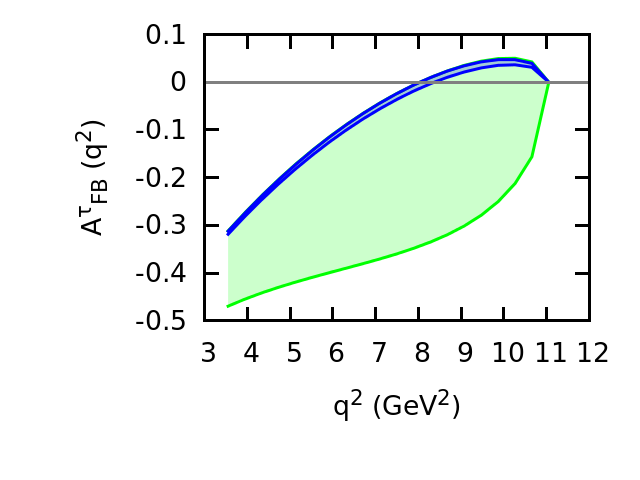}
\includegraphics[width=5.9cm,height=4.5cm]{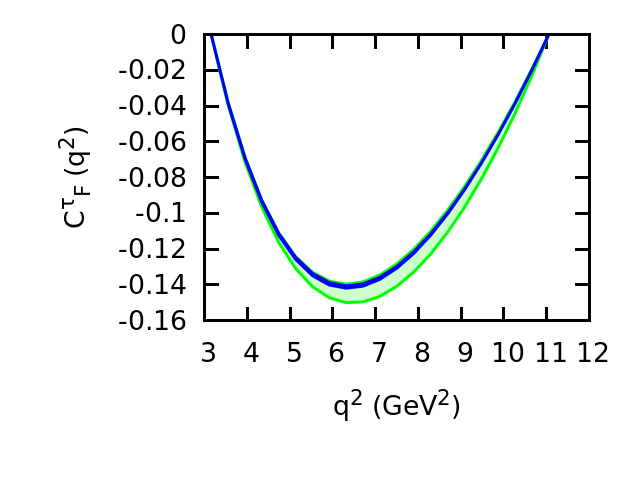}
\includegraphics[width=5.9cm,height=4.5cm]{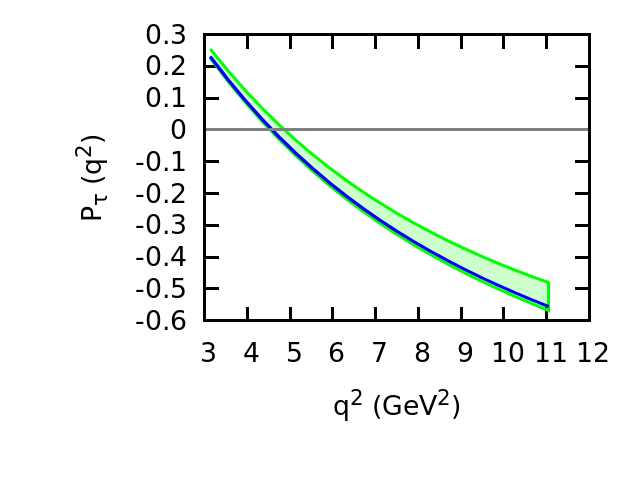}
\end{center}
\caption{The dependence of the observables DBR$(q^2)$, $R(q^2)$, $A^{\tau}_{FB}(q^2)$, $C_F^{\tau}(q^2)$, and $P_{\tau}(q^2)$ on $V_L$ and 
$V_R$ NP couplings. The allowed range in
each observable is shown in light green band once the NP couplings $(V_L,V_R)$ are varied within the allowed ranges of the left panel of 
Fig.~\ref{vlvr}. The corresponding SM prediction is shown in light blue band.}
\label{obs_vlvr}
\end{figure}

We wish to look at the effect of the new physics couplings $(V_L,\,V_R)$ on different observables such as differential branching ratio 
${\rm DBR}(q^2)$, ratio of branching ratio $R(q^2)$, forward backward asymmetry $A^{\tau}_{\rm FB}(q^2)$, the convexity parameter 
$C_F^{\tau}(q^2)$,
and the $\tau$ polarization fraction $P_{\tau}(q^2)$ for the $\Xi_b \to \Xi_c \tau\nu$ decays. In Fig.~\ref{obs_vlvr}, we show in blue the 
allowed SM bands and in green the allowed bands of each observable once the NP couplings $V_L$ and $V_R$ are switched on. It can be seen 
that once NP is included the deviation from the SM expectation is quite large in case of ${\rm DBR}(q^2)$, $R(q^2)$, and 
$A^{\tau}_{\rm FB}(q^2)$. 
However, the deviation is slightly less in case of $C_F^{\tau}(q^2)$ and $P_{\tau}(q^2)$. We observe that depending on the values of NP 
couplings,
there may or may not be a zero crossing in the forward backward asymmetry parameter $A^{\tau}_{FB}(q^2)$. In case of $P_{\tau}(q^2)$, 
the zero crossing may shift towards the higher $q^2$ value than in the SM.  

\begin{figure}[htbp]
\begin{center}
\includegraphics[width=7cm,height=6cm]{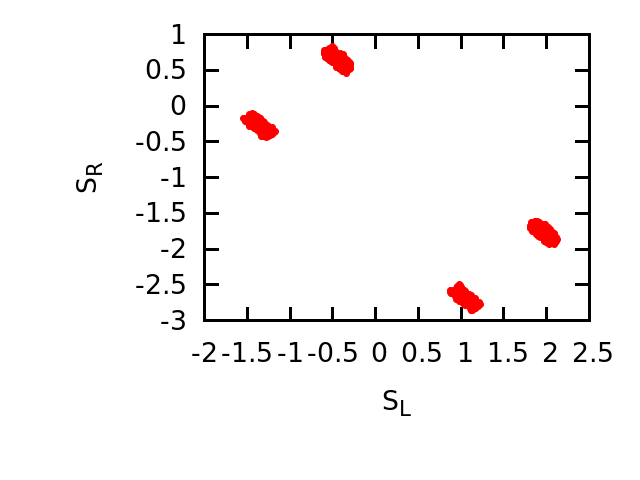}
\includegraphics[width=7cm,height=6cm]{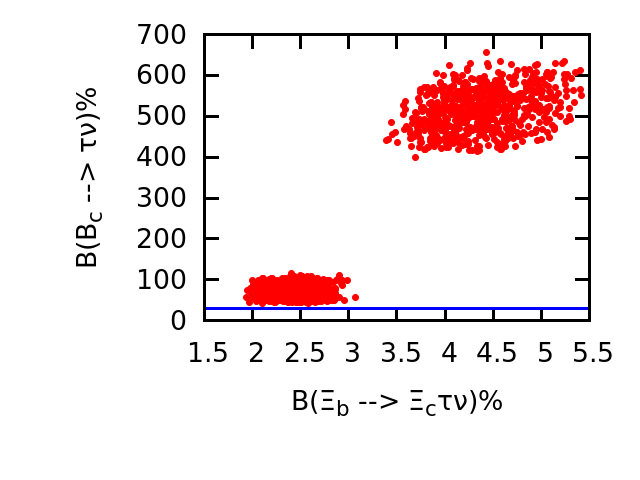}
\end{center}
\caption{Allowed regions of $S_L$ and $S_R$ obtained using the $1\sigma$ constraint coming from $R_D$ and $R_{D^{\ast}}$ are shown in the 
left panel and the corresponding ranges in $\mathcal B(B_c \to \tau\nu)$ and $\mathcal B(\Xi_b \to \Xi_c\tau\nu)$ in
the presence of these NP couplings are shown in the right panel. The blue horizontal line in the right panel corresponds to 
$\mathcal B(B_c \to \tau\nu)=30\%$.}
\label{slsr}
\end{figure}

In the second scenario, we assume that NP is coming from new scalar type of interactions, i.e, from $S_L$ and $S_R$ only.
To explore the effect of NP coming from $S_L$ and $S_R$, we vary $S_L$ and $S_R$ and impose $1\sigma$ constraint coming from the  
measured values of $R_D$ and $R_{D^{\ast}}$. The resulting ranges in $S_L$ and $S_R$ obtained using the $1\sigma$ experimental constraint are 
shown in the left panel of Fig.~\ref{slsr}. In the right panel of Fig.~\ref{slsr}, the allowed ranges in $\mathcal B(B_c \to \tau\nu)$ and 
$\mathcal B(\Xi_b \to \Xi_c\,\tau\nu)$ are shown. We see that the branching ratio of $B_c \to \tau\nu$ obtained in this scenario is rather
large, more than $30\%$. Even if we assume that $\mathcal B(B_c \to \tau\nu)$ can not be greater than $30\%$, then although $S_L$ and $S_R$
NP couplings can explain the anomalies present in $R_D$ and $R_{D^{\ast}}$, it, however, can not accommodate $B_c \to \tau\nu$ data. 
The allowed ranges in all the observables are reported in Table.~\ref{tab5}. We see a significant deviation of all the observables from the
SM prediction. It should be noted that the deviation observed in this scenario is more pronounced than the deviation observed with $V_L$ and
$V_R$ NP couplings.
\begin{table}
\begin{center}
\begin{tabular}{|c|c|c|c|c|}
\hline
$\mathcal B(\Xi_b \to \Xi_c\tau\nu)$ & $R_{\Xi_c}$ & $<A_{FB}^{\tau}>$ & $<C_F^{\tau}>$ & $<P_{\tau}>$ \\[0.2cm]
\hline
\hline
$[1.94,\,5.42]$ & $[0.200,\,0.616]$ & $[-0.179,\,0.226]$ & $[-0.052,\,-0.111]$ & $[-0.371,\,0.338]$\\[0.2cm]
\hline
\end{tabular}
\end{center}
\caption{Ranges of various observables with $S_L$ and $S_R$ NP couplings.}
\label{tab5}
\end{table}

\begin{figure}[htbp]
\begin{center}
\includegraphics[width=5.9cm,height=4.5cm]{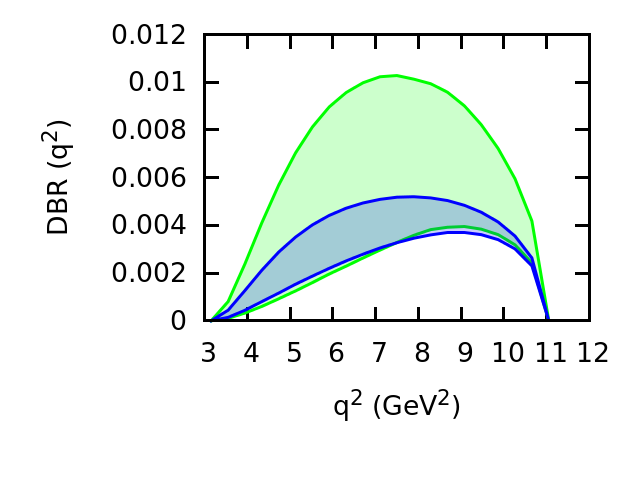}
\includegraphics[width=5.9cm,height=4.5cm]{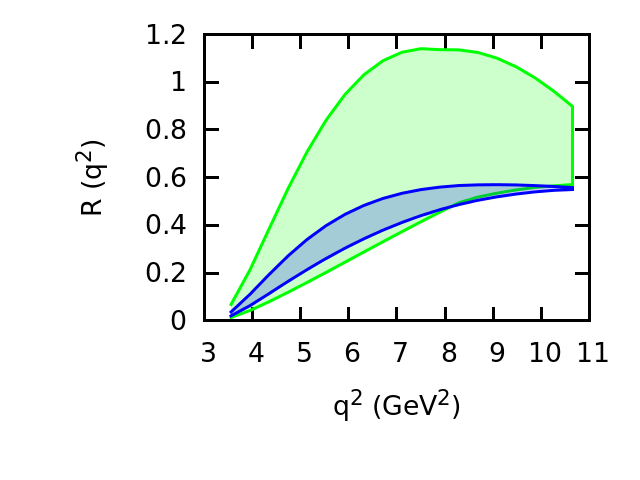}
\includegraphics[width=5.9cm,height=4.5cm]{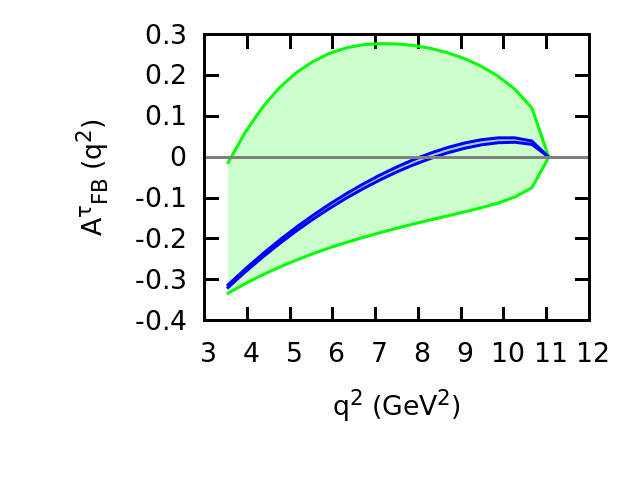}
\includegraphics[width=5.9cm,height=4.5cm]{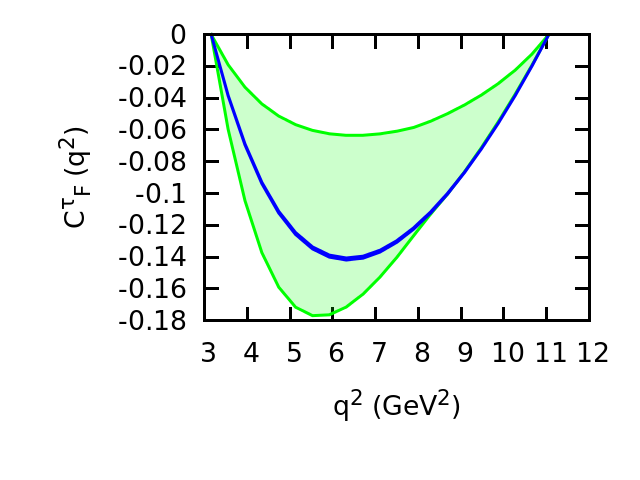}
\includegraphics[width=5.9cm,height=4.5cm]{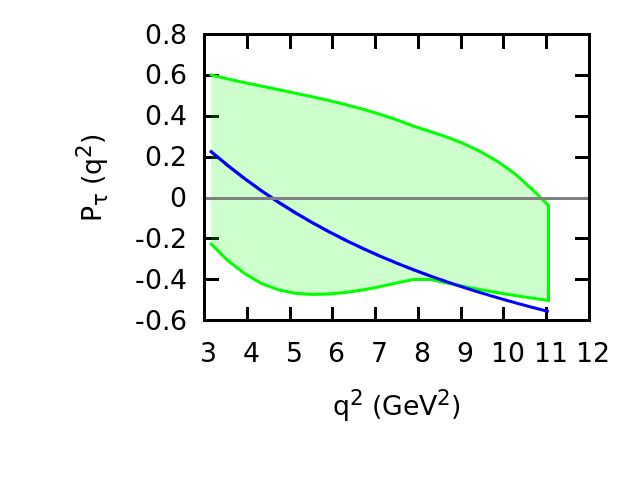}
\end{center}
\caption{The dependence of the observables DBR$(q^2)$, $R(q^2)$, $A^{\tau}_{FB}(q^2)$, $C_F^{\tau}(q^2)$, and $P_{\tau}(q^2)$ on 
$S_L$ and $S_R$ NP couplings. The allowed range in
each observable is shown in light green band once the NP couplings $(S_L,\,S_R)$ are varied within the allowed ranges shown in the left 
panel of Fig.~\ref{slsr}.
The corresponding SM prediction is shown in light blue band.}
\label{obs_slsr}
\end{figure}

We want to see the effect of these NP couplings on various $q^2$ dependent observables. In Fig.~\ref{obs_slsr}, we show how the observables 
${\rm DBR}(q^2)$, $R(q^2)$, $A^{\tau}_{\rm FB}(q^2)$, $C_F^{\tau}(q^2)$, and $P_{\tau}(q^2)$ behave as a function of $q^2$ with and without 
$S_L$ and $S_R$ NP couplings. The light blue band corresponds to the SM range whereas, the light green band corresponds to the range of the
observable with $S_L$ and $S_R$ NP couplings.
The deviations from the SM expectation is prominent in case of each observables. It should be mentioned that the deviation observed in
this scenario is more pronounced than the deviation observed with $V_L$ and $V_R$ NP couplings. Depending on the values of $S_L$ and $S_R$
NP couplings, the zero crossing point of $A_{FB}^{\tau}(q^2)$ and $P_{\tau}(q^2)$ can be quite different from the SM prediction. 

\section{Summary and conclusion}
\label{con}
Lepton flavor universality violation has been reported in various semileptonic $B$ meson decays. Tensions between SM prediction  and 
experiments exist in various semileptonic $B$ meson decays mediated via $b \to c$ charged current interactions and $b \to s\,l\,\bar{l}$
neutral current interactions. 
Study of $\Xi_b \to \Xi_c\,\tau\,\nu$ decays is important mainly for two reasons. First, it can act as
a complimentary decay channel to $B \to (D,\,D^{\ast})\tau\nu$ decays mediated via $b \to c$ charged current interactions and, in principle, 
can provide new insights into the $R_D$ and $R_{D^{\ast}}$ anomaly. Second, precise determination of the branching fractions of this decay 
modes will allow an accurate determination of the CKM matrix element $|V_{cb}|$ with less theoretical uncertainty. 

We have used the helicity formalism to study the $\Xi_b \to \Xi_c\,l\,\nu$ within the context of an effective Lagrangian in the presence of 
NP. We have defined various observables and provide predictions using form factors obtained in relativistic quark model. We have given the 
first prediction of various observables such as $R_{\Xi_c}$, $A^l_{FB}$, $P_l$, and $C_{F}^l$ for this decay mode. We also see the NP effects
on various observables for this decay mode. Let us now summarize our main results.

We first report the central values and the $1\sigma$ ranges of all the observables for the $\Xi_b \to \Xi_c\,l\,\nu$ decays within the SM. 
The SM branching ratio of $\Xi_b \to \Xi_c\,l\,\nu$ decays is at the order of $10^{-2}$.  
We observe that the integrated quantities--- forward backward asymmetry~$<A^l_{FB}>$, longitudinal 
polarization fraction of lepton~$<P_l>$, the convexity parameter~$<C_F^l>$ change considerably while going from $e$ to the $\tau$ modes. 
There is even a sign change in case of the forward backward asymmetry parameter $<A^l_{FB}>$.

For the NP analysis, we include vector and scalar type of NP interactions and explore two different NP scenarios.
In the first scenario, we consider only vector type of NP interactions, i.e, we consider that only $V_L$ and $V_R$ contributes to the
decay mode. In the second scenario, we assume that NP is coming only from scalar type of interactions, i.e, from $S_L$ and $S_R$ only.
The allowed ranges in the NP couplings are obtained by using $1\sigma$ constraint coming from the measured values of $R_D$ and
$R_{D^{\ast}}$. We also study the effect of these NP couplings on various $q^2$ dependent observables such as ${\rm DBR}(q^2)$, $R(q^2)$, 
$A^{\tau}_{\rm FB}(q^2)$, $C_F^{\tau}(q^2)$, and $P_{\tau}(q^2)$. We find significant deviations from the SM prediction once the NP couplings 
are included. However, the deviation from the SM prediction is more pronounced in case of scalar NP interactions $S_L$ and $S_R$. It should be
mentioned that $\mathcal B(B_c \to \tau\nu)$ put a severe constraint on $S_L$ and $S_R$ NP couplings. However, the allowed range obtained for
$\mathcal B(B_c \to \tau\nu)$ with $V_L$ and $V_R$ NP couplings is consistent with the $\mathcal B(B_c \to \tau\nu)\le 5\%$ obtained in the SM.

Although, there is hint of NP in the meson sector, NP is not yet established. Study of $\Xi_b \to \Xi_c\,l\,\nu$ decays both theoretically
and experimentally is well motivated because of the longstanding anomalies present in $R_D$ and $R_{D^{\ast}}$. It would be interesting to 
find out similar hint of NP in the semileptonic baryonic decays as 
well. At the same time, a precise measurement of $\mathcal B(\Xi_b \to \Xi_c\,l\,\nu)$ and a precise determination of $\Xi_b \to \Xi_c$
transition form factors will allow an accurate determination of the CKM matrix element $|V_{cb}|$.

\end{document}